\documentstyle[epsfig,epsf]{article}
\begin{document}

\begin{center}
{\Large {\bf Search for massive rare particles with the SLIM experiment}}
\end{center}

\vskip .7 cm

\begin{center}
MIRIAM GIORGINI on behalf of the SLIM Collaboration{\footnote {The SLIM 
Collaboration (Bolivia, Canada, Italy, Pakistan): S.Balestra, S. Cecchini, F. 
Fabbri, G. Giacomelli, M. Giorgini, A. Kumar, S. Manzoor, J. McDonald, E. 
Medinaceli, J. Nogales, L. Patrizii, J. Pinfold, V. Popa, O. Saavedra, G. 
Sher, M. Shahzad, M. Spurio, V. Togo, A. Velarde and A. Zanini.}}
 
 \par~\par
{\it Dept of Physics, Univ. of Bologna and INFN, \\
V.le C. Berti Pichat 6/2, Bologna, I-40127, Italy\\} 

E-mail: miriam.giorgini@bo.infn.it

\par~\par

Talk given at the 10$^{th}$ ICATPP Conference on
Astroparticle, Particle, Space Physics, Detectors and Medical Physics 
Applications, Como, Italy, 8-12 October 2007.

\vskip .7 cm
{\large \bf Abstract}\par
\end{center}

{\normalsize 
The SLIM experiment is a large array of
nuclear track detectors located at the Chacaltaya High Altitude
Laboratory (5260 m a.s.l.). The preliminary results from the analysis
of $\sim 383$ m$^2$ exposed for 4.07 y are here
reported. The detector is sensitive to Intermediate Mass Magnetic
Monopoles, $10^5 < M_M < 10^{12}$ GeV, and to SQM nuggets and
Q-balls, which are possible Dark Matter candidates.}

%\vspace{5mm}

\section{Introduction}\label{sec:intro}

Grand Unified Theories (GUT) of the strong and electroweak 
interactions predict the existence of magnetic monopoles (MMs), produced  
in the early Universe at the end of the GUT epoch, with 
very large masses, $M_M > 10^{16}$ GeV. GUT poles in the cosmic 
radiation should be 
characterized by low velocity and 
relatively large energy losses \cite{MMs}. At present the MACRO
experiment has set the best limit on GUT MMs for 
$4 \cdot 10^{-5} < \beta < 0.5$ \cite{MACRO}.

Intermediate Mass Monopoles (IMMs) [$10^5 \div  10^{12}$ GeV] 
could also be present in the cosmic radiation; they may have been
produced in later phase transitions in the early Universe
\cite{IMMs}. The recent interest in IMMs is also
connected with the possibility that they could yield the highest
energy cosmic rays \cite{UHECR}. IMMs may have
relativistic velocities since they could be accelerated  
in one coherent domain of the galactic magnetic field. In
this case one would have to look for downgoing fast ($\beta > 0.1$)
heavily ionizing MMs.

Besides MMs, other massive particles have been hypothesized to exist 
in the cosmic radiation and to be components of the galactic
cold dark matter: nuggets of Strange Quark Matter (SQM), called
nuclearites when neutralized by captured electrons, and Q-balls. SQM
consists of aggregates of u, d and s quarks (in approximately 
equal proportions) with slightly positive electric charge \cite{nuclr}. It 
was suggested that SQM may be
the ground state of QCD. They should be stable for all baryon
numbers in the range between ordinary heavy nuclei and neutron stars
(A $\sim 10^{57}$). 
 Nuclearite interaction with matter depend on their 
mass and size. In ref. \cite{SLIM05/5} different mechanisms of energy loss
and propagation in relation to their detectability with the SLIM
apparatus are considered. In the absence of any
candidate, SLIM will be able to rule out some of the hypothesized
propagation mechanisms. 

Q-balls are super-symmetric coherent states of
squarks, sleptons and Higgs fields, predicted by minimal
super-symmetric generalizations of the Standard Model \cite{qballs}.
They could have been produced in the early Universe. Charged Q-balls
should interact with matter in ways not too dissimilar from those of
nuclearites.
  
After a short description of the apparatus, we
present the calibrations, the analysis procedures and the results from
the SLIM experiment.

\section{Experimental procedure}
The SLIM (Search for LIght magnetic Monopoles) experiment, based on 440
m$^2$ of Nuclear Track Detectors (NTDs), was deployed at the
Chacaltaya High Altitude 
Laboratory (Bolivia, 5260 m a.s.l.) since 2001 \cite{slim}. 
 The air temperature is recorded 3 times a day. From the observed ranges 
of temperatures we conclude that no
significant time variations occurred in the detector
response. The radon activity and the flux of cosmic ray neutrons were 
measured by us and by other authors \cite{neutron}.  
Another 100 m$^2$ of NTDs were installed at Koksil 
(Pakistan, 4600 m a.s.l.) since 2003.

Extensive test studies were made in order to improve the etching
procedures of CR39 and Makrofol, improve the scanning and
analysis procedures and speed, and keep a good scan efficiency.  
``Strong'' and ``soft'' etching conditions have been defined
\cite{calib}. Strong etching conditions (8N KOH + 1.25\%
Ethyl alcohol at 77 $^\circ$C for 30 hours) are used for the first
CR39 sheet in each module, in order to produce large tracks, easier to
detect during scanning. Soft etching conditions (6N NaOH +
1\% Ethyl alcohol at 70 $^\circ$C for 40 hours) are applied to the other
CR39 layers in a module, if a candidate track is found in the first
layer. It allows more reliable measurements of the Restricted Energy
Loss (REL) and of the direction of the incident particle. Makrofol
layers are etched in 6N KOH + Ethyl alcohol (20\% by volume), at
50 $^\circ$C. 

\begin{figure}
\begin{center}
\psfig{file=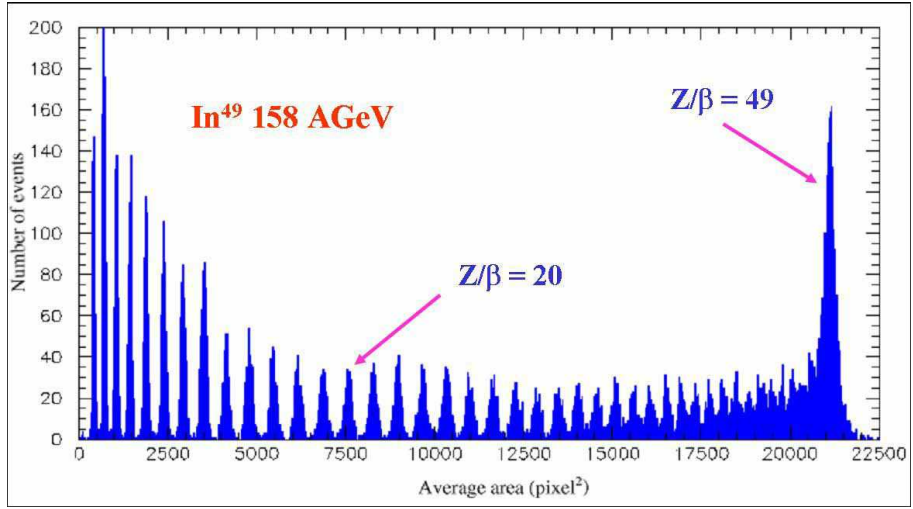,width=12cm}
\end{center}
\caption{Calibrations of CR39 nuclear track detectors
 with 158 AGeV In$^{49+}$ ions and their fragments.}
\label{fig:fig1}
\end{figure}
  
The detectors have been calibrated using 158 AGeV In$^{49+}$ (see
Fig. \ref{fig:fig1}) and 30 AGeV Pb$^{82+}$ beams. For soft etching 
conditions the threshold in CR39 is
at REL $\sim$ 50 MeV cm$^2$ g$^{-1}$; for strong etching the threshold
is at  REL $\sim$ 250 MeV cm$^2$ g$^{-1}$.  
Makrofol has a higher threshold (REL $\sim 2.5$ GeV
cm$^2$ g$^{-1}$) \cite{makrofol}. The CR39 allows the detection 
of IMMs  with two units
Dirac charge in the whole $\beta$-range of $4 \cdot 10^{-5} < \beta <
1$. The Makrofol is useful for the detection of fast MMs; nuclearites 
with $\beta \sim 10^{-3}$ can be detected by both CR39 and Makrofol.

The analysis of a SLIM module starts by etching the top CR39
sheet using  strong conditions, reducing its thickness from 1.4 mm
to $\sim 0.6$ mm. Since MMs, 
nuclearites and Q-balls should have a constant REL through the stack,
the signal looked for is a hole or a biconical track with the
two base-cone areas equal within the experimental uncertainties. The
sheets are scanned with a low magnification stereo
microscope. Possible candidates are further analyzed with a high 
magnification microscope. The size of surface tracks is measured on
both sides of the sheet.  We require the two values to be equal within
3 times the standard deviation of their difference. A track is
defined as  a ``candidate'' if the REL and the incidence angles on the
front and back sides are equal to within 15\%.
To confirm the candidate track, the bottom CR39 layer is then
etched in soft conditions; an accurate scan under an
optical microscope with high magnification is performed  in a
region of about 0.5 mm around the expected candidate
position. If a two-fold coincidence is found the middle layer
of the CR39 (and in case of high Z candidate, the Makrofol layer) is
analyzed with soft conditions.

\begin{figure}
\begin{center}
\mbox{\psfig{file=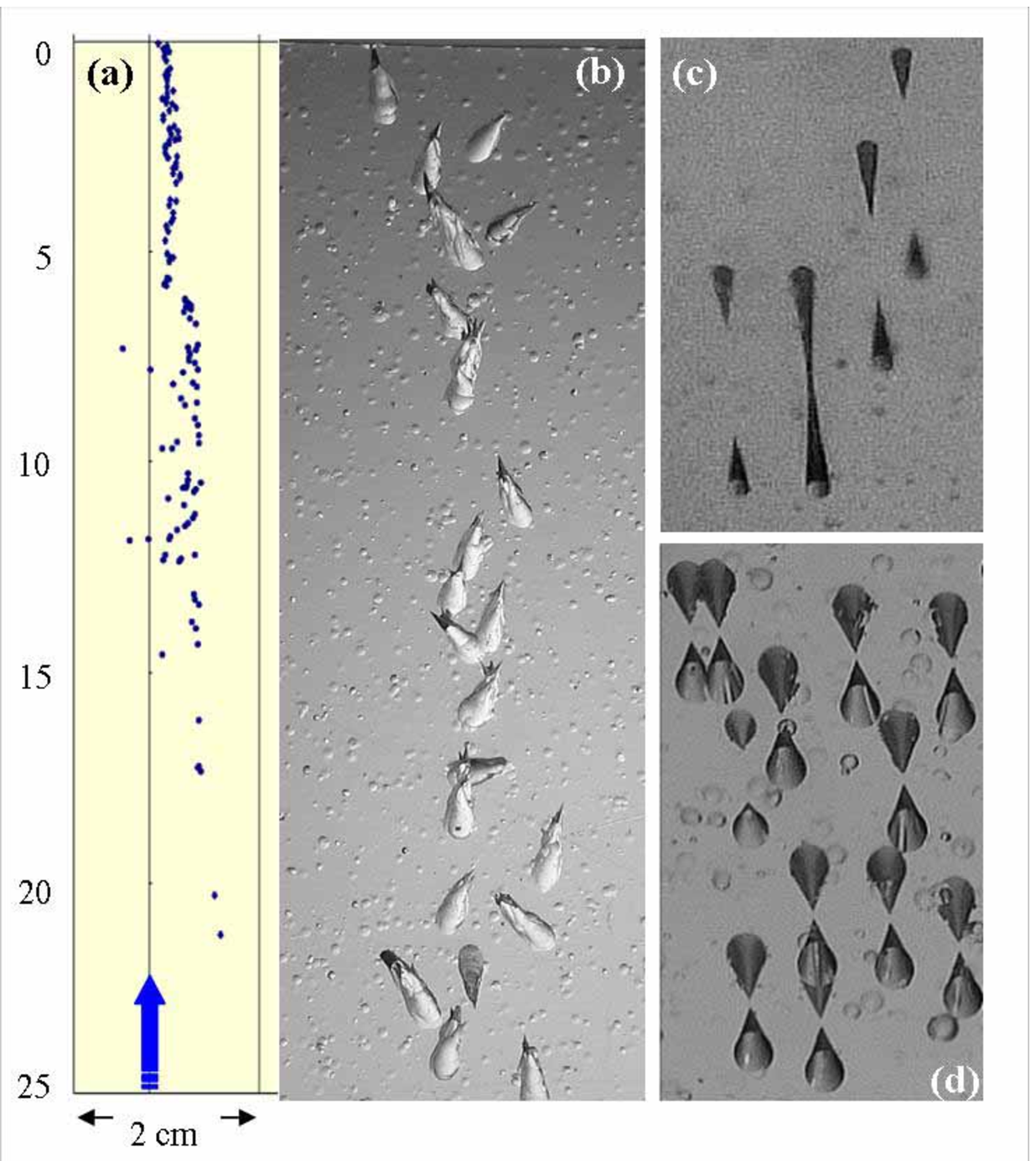,width=7cm}}
\hspace{5mm}
\mbox{\psfig{file=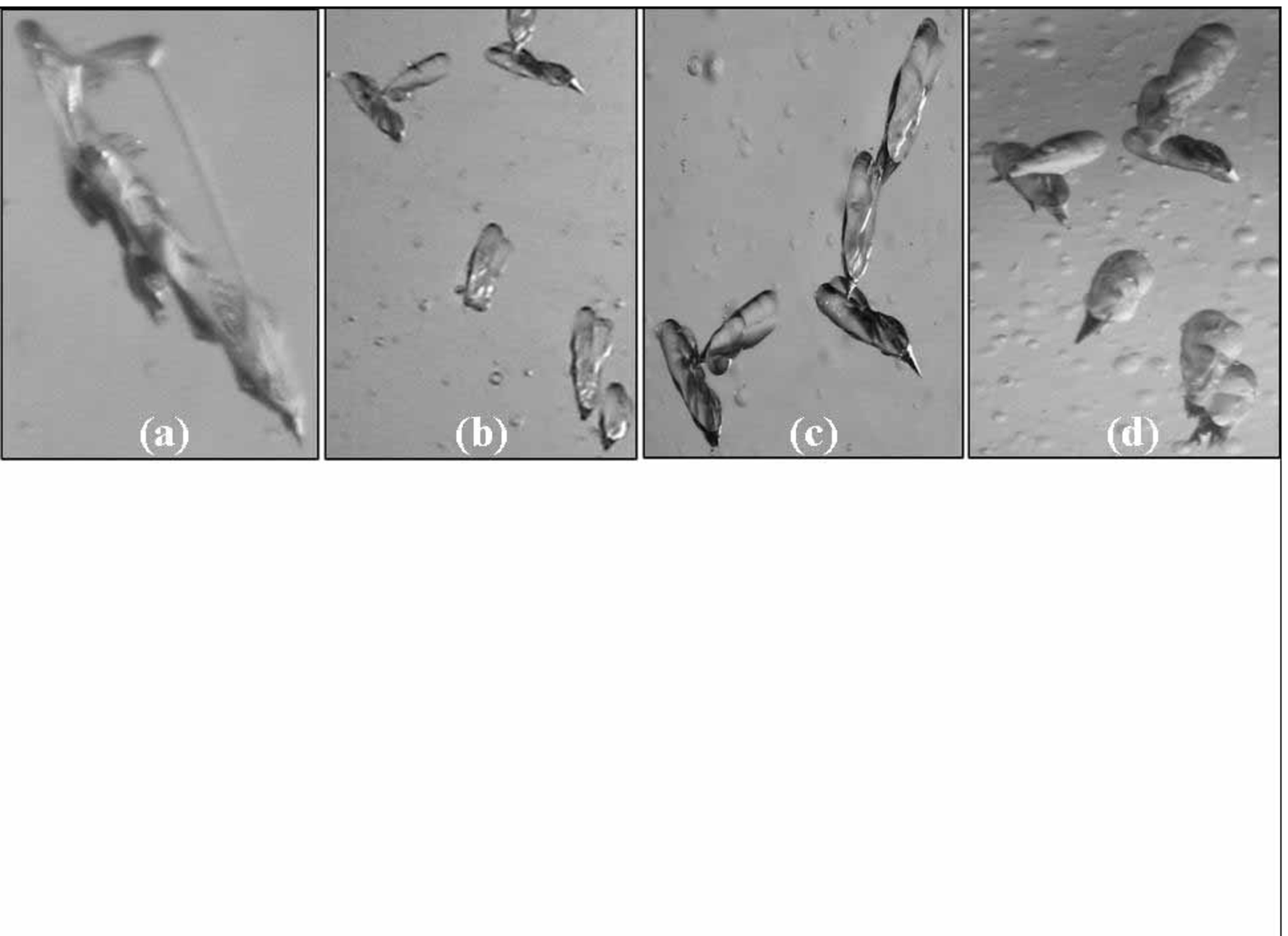,width=8.5cm}}
\end{center}
\caption{Left: (a) Global view of the ``event'' tracks in
  the L1 layer of wagon 7408. (b) Microphotographs of the
  22 cones at the top of Fig. 3a. (c) Normal tracks of 158 AGeV
  Pb$^{82+}$ ions and their fragments (soft
  etching), and (d) of 400 AMeV Fe$^{26+}$ ions and their fragments
  (strong etching).
Right: Example of ``tracks'' in the L6 layer of wagon 7410:
(a) after 30h of soft etching, (b) after 5h more of strong etching, (c)
 after 4h more of strong etching and (d) after 10h more of strong etching.}
\label{fig:fig2}
\end{figure}

\section{Non reproducible candidates}
\label{sec:nonrep}
In 2006 the SLIM experiment found a very strange event when analyzing 
the top CR39 layer of stack 7408. We found a sequence
of many ``tracks'' along a 20 cm 
line; each of them looked complicated and very
different from usual ion tracks, see Fig. \ref{fig:fig2}left(a,b). For
comparison Fig. \ref{fig:fig2}left(c) shows ``normal'' tracks from 158 AGeV
Pb$^{82+}$  ions and their fragments and Fig. \ref{fig:fig2}left(d) shows 
tracks from 400 AMeV Fe$^{26+}$ ions.

Since that ``event'' was rather
peculiar, we made a detailed study of all the sheets of
module 7408, and a search for similar events and
in general for background tracks in all NTD sheets in the wagons
around module 7408 (within a $\sim$ 1 m distance from module 7408). We 
etched
``softly'' all the sheets in order to be able to follow the evolution of
the etch-pits. A second event was found in the CR39 bottom layer (top
face) of module 7410, see Fig. \ref{fig:fig2}right.  
Some background tracks in other modules were found after 30 h of
soft etching. We decided to further
etch ``strongly'' the 7410-L6 layer 
in short time steps (5h) and to follow the evolution of the
``tracks'' by systematically making photographs
at each etching step. After additional strong etching,
 the ``tracks'' began more and more similar to those in the 7408-L1
layer, see Fig. \ref{fig:fig2}right(b,c,d). The presence of this second
event/background and its evolution with 
increasing etching casts stronger doubts on the
event interpretation and supports a ``background'' interpretation also
of the ``tracks'' in module 7408. We made different hypotheses and we
checked them with the Intercast Co. Since 1980
we analyzed more than 1000 m$^2$ of CR39 using different etching
conditions and we have not seen before any of the above mentioned
cases. It appears that we may have been hit by an extremely rare
manufacturing defect involving 1 m$^2$ of CR39.

\begin{figure}
\begin{center}
\psfig{file=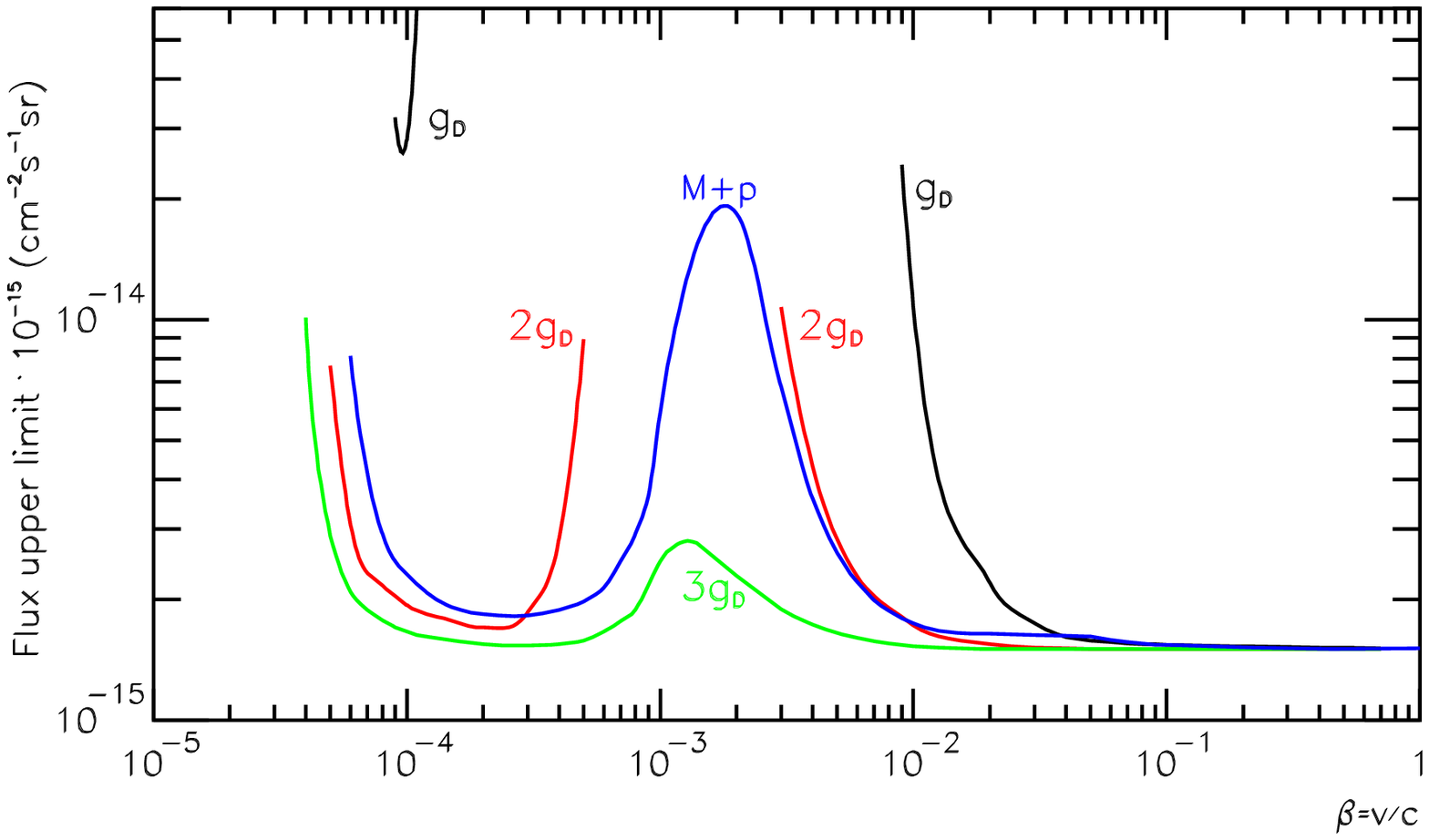,height=6.8cm,width=8.3cm}
\hspace{-5mm}
\psfig{file=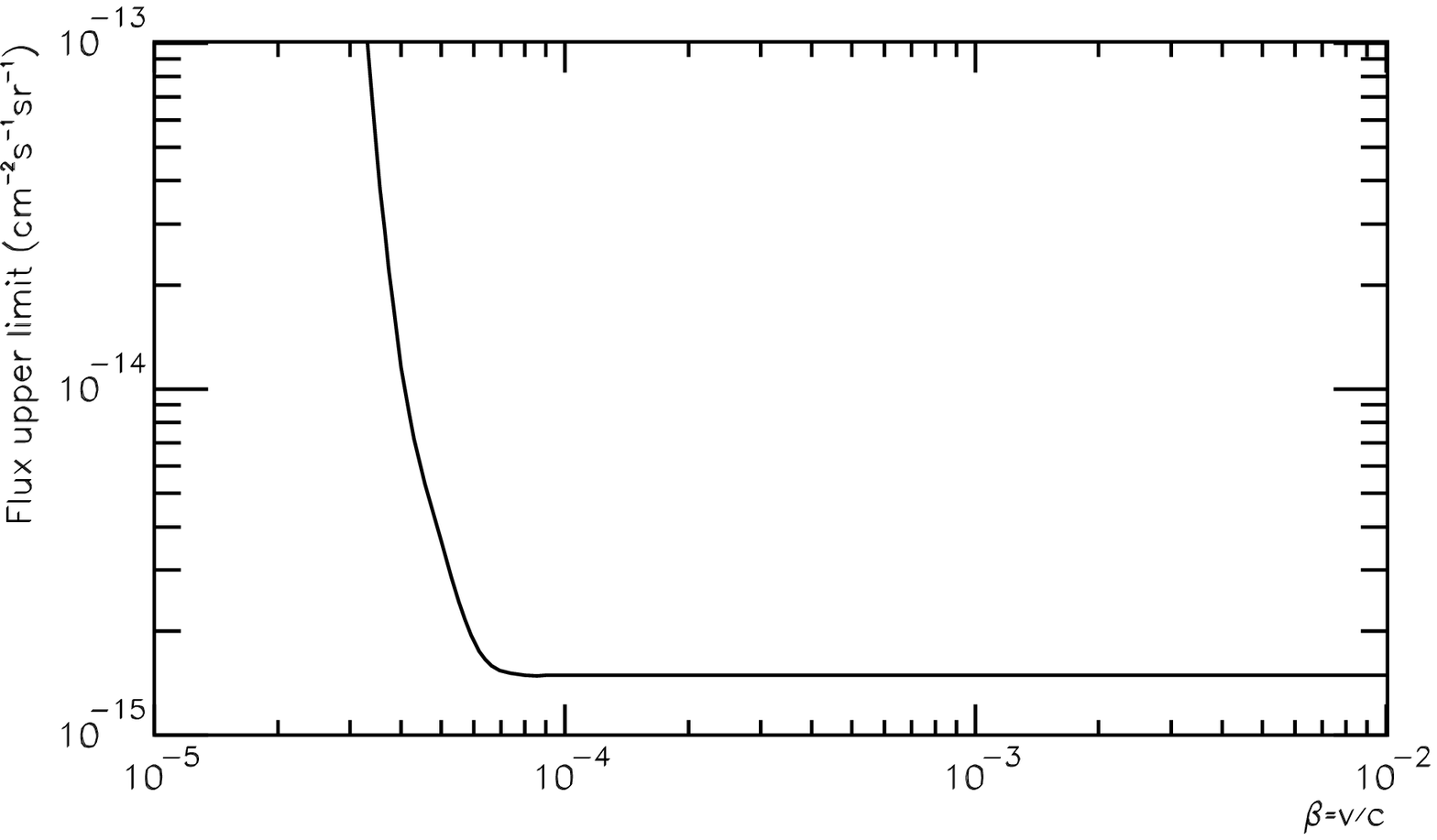,height=6.8cm,width=8.3cm}
\end{center}
\caption{Left: 90\% C.L. upper limits for a downgoing
 flux of IMMs with $g = g_D,~2g_D,~3g_D$ and for dyons
 (M+p) plotted vs $\beta$. Right: 90\% C.L. upper limits for a downgoing
 flux of nuclearites with $M_N \leq 8.4 \cdot 10^{14}$ GeV.}
\label{fig:fig4}
\end{figure}

\section{Results and Conclusions}
We etched and analyzed 383 m$^2$ of CR39, with an average exposure
time of $\sim 4.07$ years. No candidate passed the search criteria: the 90\%
C.L. upper limits for a downgoing flux of IMMs with $g = g_D,~2g_D,~3g_D$ 
and for dyons (M+p)
are at the level of $\sim 1.5 \cdot 10^{-15}$ cm$^{-2}$ s$^{-1}$
sr$^{-1}$ for $\beta \geq 4 \cdot 10^{-2}$, see Fig. \ref{fig:fig4}left.
The same sensitivity was reached also for nuclearites with 
$\beta \geq 10^{-4}$ (Fig. 
\ref{fig:fig4}right) and Q-balls coming from above with galactic velocities.

\end{document}